\documentclass[12pt,english]{article}
\usepackage[T1]{fontenc}
\usepackage[latin9]{inputenc}
\usepackage{array}
\usepackage{longtable}
\usepackage{booktabs}
\usepackage{multirow}
\usepackage{amsbsy}
\usepackage{graphicx}
\usepackage{setspace}
\makeatletter

\providecommand{\tabularnewline}{\\}

\usepackage{aStyle}

\makeatother

\usepackage{babel}
\begin{document}
\begin{doublespace}
\begin{center}
\textbf{\Large{}Energy Minimization in Fluid Flow through Tubes
and Networks of Various Geometries}\vspace{-1.3cm}
\par\end{center}
\end{doublespace}

\begin{center}
Taha Sochi\footnote{University College London - Department of Physics \& Astronomy - Gower
Street - London - WC1E 6BT. Email: t.sochi@ucl.ac.uk.}
\par\end{center}

\noindent \textbf{Abstract}: In this paper we continue our previous
investigation about energy minimization in the flow of fluids through
tubes and networks of interconnected tubes of various geometries.
We will show that the principle of energy minimization holds independent
of the geometry of the tubes and networks of such interconnected tubes
and independent of the type of fluid in such geometries where in this
regard we consider generalized Newtonian fluids. We consider in this
investigation the flow of Newtonian fluids through tubes and networks
of interconnected tubes of elliptical, rectangular, equilateral triangular
and concentric circular annular cross sectional geometries. We also
consider a combination of geometric factor with a fluid type factor
by showing that the principle of energy minimization holds in the
flow of some non-Newtonian fluids (namely power law, Ellis and Ree-Eyring
fluids) through tubes and networks of interconnected tubes of elliptical
cross sections. The relevance of this study extends beyond tubes and
networks of fluid flow to include for instance porous media and electrical
networks.\footnote{All symbols and abbreviations are defined in $\S$ \hyperref[Nomenclature]{Nomenclature}
in the back of this paper.}\vspace{0.3cm}


\clearpage{}

\tableofcontents{}

\clearpage{}

\section{Introduction}

Optimization is one of the fundamental principles of Nature where
physical systems tend to reach their targets through optimizing (usually
minimizing) certain physical quantities and attributes. A well known
example is Fermat's principle of least time. In particular, energy
minimization is a well know principle in many physical systems such
as the tendency of excited atomic systems to decay to the lowest available
energy levels within certain rules and conditions.

There are quite few investigations on the theoretical formulation
and practical application of optimization principles as governing
rules that determine the behavior and regulate the design of fluid
dynamics systems (see for instance \cite{Murray1926a,Murray1926c,RossittiL1993a,SharmaASZ2011}).
However, we are not aware of the use of optimization principles (and
energy minimization in particular) to resolve the pressure fields
and obtain flow dynamics quantities and parameters (such as volumetric
flow rates) of fluid flow through conduits and networks of interconnected
conduits. Yes, a few years ago we have investigated some of these
issues (see \cite{SochiPresSA2014,SochiPresSA22014,SochiOptMulVar2014})
using energy minimization as a viable principle for resolving the
flow fields in conduits and networks as well as a fundamental principle
that underlies, and hence explains, the behavior of fluid dynamics
systems in general.\footnote{We should note that variational principles (which we investigated
in a number of studies; see for instance \cite{SochiVariational2013,SochiTubeCircSA2015})
are also optimization principles.}

In our previous investigations (which we cited already) about energy
minimization in the flow of fluids through tubes and networks of interconnected
tubes we established that energy minimization principle holds in both
cases (i.e. tubes and networks) regardless of the type of fluid (such
as Newtonian or power law or Ellis or Ree-Eyring or other types of
generalized Newtonian fluids). However, those investigations were
limited to tubes of circular (and simply-connected) cross sectional
geometry and networks of interconnected circular tubes.

In the present investigation we generalize those results by extending
them to some of the most common types of tubes of non-circular (and
non-simply-connected) cross sectional geometries and networks of interconnected
such tubes. So, we will extend the energy minimization principle to
tubes and networks of interconnected tubes of elliptical, rectangular,
equilateral triangular and concentric circular annular cross sectional
geometries with regard to the flow of Newtonian fluids through these
geometries. In addition to this, we will combine non-circular geometry
factor with the non-Newtonian fluid factor by extending the principle
of energy minimization to the flow of some generalized Newtonian fluids
(namely power law, Ellis and Ree-Eyring) through tubes and networks
of interconnected tubes of elliptical cross sections.

The result that we will reach in this investigation is that the principle
of energy minimization is valid in general, i.e. its validity is independent
of the type of fluid and the geometry of conduits (as well as other
attributes and characteristics of networks such as their dimensionality
and topology). The relevance of the current investigation (as well
as our previous investigations) extends beyond the flow of fluids
through tubes and networks of interconnected tubes. For example, the
principle of energy minimization should similarly apply to the flow
of fluids through porous media since porous media are essentially
networks of interconnected tubes of irregular geometry and random
topology.

In fact, the principle of energy minimization which we establish in
this investigation should extend even to networks other than fluid
transportation networks. For example, electrical networks of ohmic
conductors should also be subject to the energy minimization principle
because of the similarity in the fundamental physics between networks
of interconnected tubes (for fluid flow) and networks of interconnected
ohmic resistors (for electric current flow).

Our plan in this paper (following this introduction) is to prepare
the scene for this investigation by providing a general theoretical
background about this investigation where we mostly refer to our previous
investigations in this regard (see $\S$ \ref{secBackground}). We
then investigate energy minimization principle in the flow of Newtonian
fluids through tubes and networks of interconnected tubes of various
geometries, i.e. tubes and networks of interconnected tubes of elliptical
cross sections, rectangular cross sections, equilateral triangular
cross sections, and concentric circular annular cross sections (see
$\S$ \ref{secNewtonianGeometries}). We then investigate energy minimization
principle in the flow of three non-Newtonian fluid types (namely power
law, Ellis and Ree-Eyring) through tubes and networks of interconnected
tubes of elliptical cross sectional geometry (see $\S$ \ref{secNonNewtonian}).
The paper is finalized with a list of the main achievements and conclusions
of this investigation (see $\S$ \ref{secConclusions}).

\clearpage{}

\section{\label{secBackground}Theoretical Background and Implementation}

The theoretical background of this investigation is largely explained
in \cite{SochiPresSA2014,SochiPresSA22014,SochiOptMulVar2014} (and
\cite{SochiOptMulVar2014} in particular). So, in this section we
just give an outline or reminder of what have been given already.

\textbf{The traditional method}\footnote{We may refer to the results of traditional method by labels like ``analytical
solutions'' or ``Poiseuille-type solutions''.} for resolving the pressure and volumetric flow rate fields in fluid
conducting devices is to use the conservation principles, which are
normally based on the mass continuity and momentum conservation, in
conjunction with the constitutive relations that link the stress to
the rate of deformation and are specific to the particular types of
fluid employed to model the flow \cite{Skellandbook1967,BirdbookAH1987,Whitebook1991,PapanastasiouGABook1999}.

For single conduits, this usually results in an analytical expression
that correlates the volumetric flow rate to the applied pressure drop
as well as other dependencies on the parameters of the conduits, such
as the radius and length of the tube, and the parameters of the fluid
such as the shear viscosity and yield stress. For networks of interconnected
conduits, the analytical expression for the single conduit for the
particular fluid model can be exploited in a numeric solution scheme,
which is normally of iterative nature such as the widely used Newton-Raphson
procedure for solving a system of simultaneous non-linear equations,
in conjunction with the mass conservation principle and the given
boundary conditions to obtain the flow fields in the network (see
for instance \cite{SochiThesis2007,SochiPois1DComp2013}).

\textbf{The energy minimization method} for resolving the pressure
and volumetric flow rate fields in fluid conducting devices (which
is the foundation of the present investigation as a continuation of
the previous investigations of \cite{SochiPresSA2014,SochiPresSA22014,SochiOptMulVar2014})
employs a different strategy which will be outlined in the following
paragraphs.

The time rate of energy consumption, $I$, for transporting a certain
quantity of fluid through a single conducting device, considering
the relevant flow assumptions and conditions, is given by:

\begin{equation}
I=\Delta p\,Q
\end{equation}
where $\Delta p$ is the pressure drop across the conducting device
and $Q$ is the volumetric flow rate of the transported fluid through
the device. For a flow conducting device that consists of or discretized
into $m$ conducting elements indexed by $l$, the total energy consumption
rate, $I_{t}$, is given by:

\begin{equation}
I_{t}(p_{1},\ldots,p_{N})=\sum_{l=1}^{m}\Delta p_{l}Q_{l}\label{ItEq}
\end{equation}
where $N$ is the number of the boundary and internal nodes. For a
single duct, the conducting elements are the discretized sections,
while for a network they represent the conducting ducts as well as
their discretized sections if discretization is employed.

Starting from randomly selected values for the internal nodal pressure,
with the given pressure values for the inlet and outlet boundary nodes,
the role of the global multi-variable optimization algorithms in the
above-outlined energy consumption model is to minimize the cost function,
which is the time rate of the total energy consumption for fluid transportation
$I_{t}$ as given by Eq. \ref{ItEq}, by varying the values of the
internal nodal pressure while holding the pressure values at the inlet
and outlet boundary nodes as constants. The volumetric flow rates,
$Q$, that have to be used in Eq. \ref{ItEq} for the employed fluid
models are given by the expressions in Table \ref{QGTable} for Newtonian
fluids and Table \ref{QETable} for non-Newtonian fluids (as will
be explained next in $\S$ \ref{secNewtonianGeometries} and $\S$
\ref{secNonNewtonian}).

This energy minimization method was implemented using three deterministic
global multi-variable optimization algorithms and one stochastic.
The deterministic algorithms are: Conjugate Gradient, Nelder-Mead,
and Quasi-Newton, while the stochastic is the Global algorithm of
Boender \etal. For more details about the three employed deterministic
algorithms we refer to standard textbooks that discuss these algorithms,
while for the Stochastic Global algorithm we refer to \cite{BoenderKTS1982}.\footnote{Also see: http://jblevins.org/mirror/amiller/global.txt web page.}

As for single tubes, it is a special case of the forthcoming 1D linear
networks of serially connected pipes (see for instance Tables \ref{TabNewtEllipseFig}
and \ref{TabPLEllipseFig}) where all the pipes in the ensemble have
the same parameters (such as the semi-major and semi-minor axes for
tubes of elliptical cross sections). In this regard all the implemented
optimization algorithms produced results which are virtually identical
to the analytical solutions for the certain types of fluid as given
in Tables \ref{QGTable} and \ref{QETable}.

Regarding the networks, due to the difficulty of presenting the results
graphically for the two-dimensional and three-dimensional networks,
we present in Figures \ref{NewtEllipseFig}-\ref{NewtAnnFig} and
Figures \ref{PLEllipseFig}-\ref{ReeEllipseFig} a sample of the results
obtained from a range of 1D linear networks (where all the produced
results, as we see in the sample, of the traditional method and the
energy minimization method are virtually identical). Similar results
were obtained from representative samples of 2D and 3D networks of
various sizes and topologies (similar to what we did in \cite{SochiOptMulVar2014}).

\clearpage{}

\section{\label{secNewtonianGeometries}Newtonian Flow in Various Geometries}

\begin{figure}
\centering\includegraphics[scale=0.8]{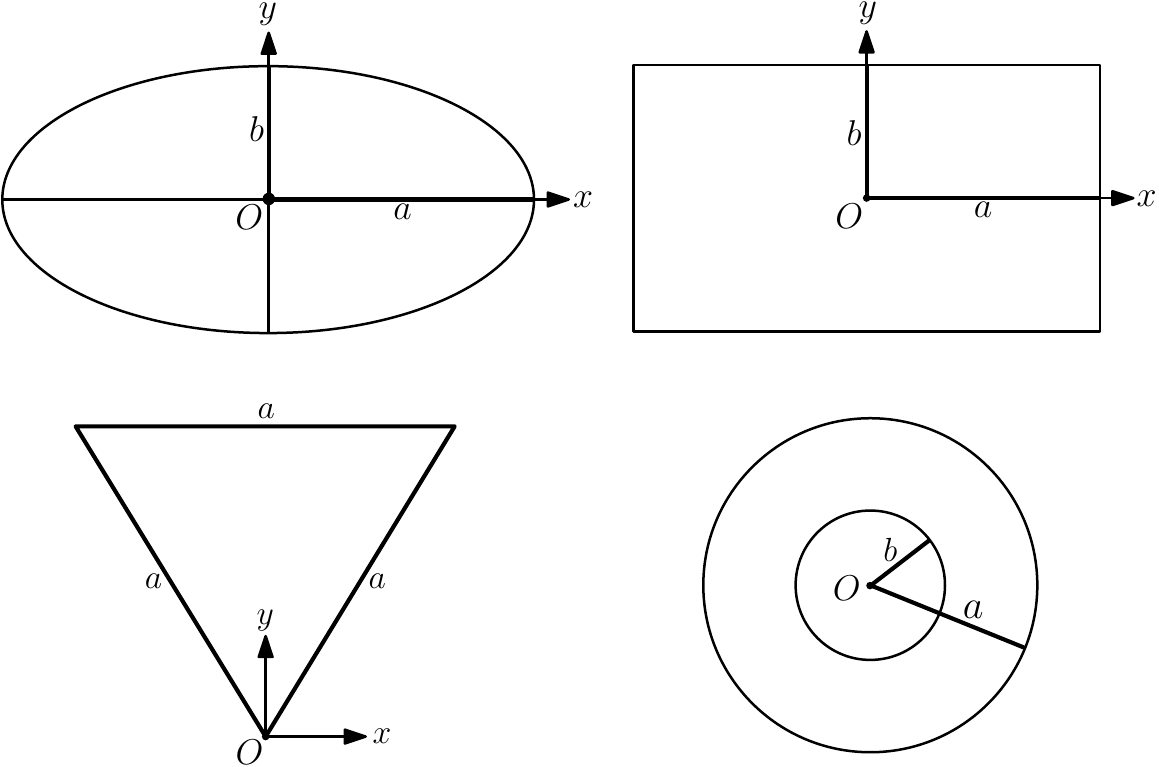}\label{FigShapes}\caption{The settings of the elliptical, rectangular, equilateral triangular,
and concentric circular annular cross sections of the tubes where
$O$ is the origin of coordinates. The $z$ axis is emanating from
the origin and is perpendicular to the plane of cross sections. See
Table \ref{QGTable} and Table \ref{QETable}.}
\end{figure}

In this section we outline our investigation of the flow of Newtonian
fluids through tubes and networks of interconnected tubes of elliptical
cross sections, rectangular cross sections, equilateral triangular
cross sections, and concentric circular annular cross sections (see
Figure \ref{FigShapes}). In this regard we use the analytical expressions
of Table \ref{QGTable} (as given for instance in \cite{Whitebook1991,Lekner2007})
for the volumetric flow rate of Newtonian fluid flow through these
geometries where the coordinate and parametric settings of these geometries
with regard to these expressions are illustrated in Figure \ref{FigShapes}.

We conducted thorough investigations and computations of Newtonian
flow through tubes and networks of various dimensionality (i.e. 1D,
2D and 3D) and topology (such as linear, fractals, and irregulars
based on cubic and orthorhombic lattices) and of various cross sectional
geometries (i.e. elliptical, rectangular, equilateral triangular and
concentric circular annular) using Newtonian fluids of various viscosity.
All these investigations and computations lead to the definite conclusion
that energy minimization holds in all these cases regardless of the
type of vessel (i.e. whether single tube or network of interconnected
tubes) and its geometry, regardless of the dimensionality and topology
of network, and regardless of the viscosity of fluid.

A sample of these investigations and computations for some 1D linear
networks (see Tables \ref{TabNewtEllipseFig}-\ref{TabNewtAnnFig})
are given in Figures \ref{NewtEllipseFig}-\ref{NewtAnnFig}, i.e.
Figure \ref{NewtEllipseFig} for the flow of a Newtonian fluid through
elliptical geometry, Figure \ref{NewtRectFig} for the flow of a Newtonian
fluid through rectangular geometry, Figure \ref{NewtTriangFig} for
the flow of a Newtonian fluid through equilateral triangular geometry,
and Figure \ref{NewtAnnFig} for the flow of a Newtonian fluid through
concentric circular annular geometry. The use of 1D linear networks
in this sample demonstration is because of the ease of displaying
and clarity of showing and interpreting the results of this type of
networks which can be easily displayed on 2D plots (as seen in Figures
\ref{NewtEllipseFig}-\ref{NewtAnnFig}) and hence they are clearly
analyzed and interpreted. 

As we see, this representative sample shows that the solutions that
we obtained from the employment of the principle of energy minimization
are virtually identical to the Poiseuille-type analytical solutions
(as outlined in $\S$ \ref{secBackground}). So, we can conclude with
certainty that the principle of energy minimization holds in general
(i.e. independent of the type of vessel and its geometry, the dimensionality
and topology of network, and the type of Newtonian fluid).

\clearpage{}

\begin{table}
\begin{centering}
\caption{The four cross sectional geometries of tubes and their corresponding
volumetric flow rates ($Q$) for the flow of Newtonian fluids through
these tubes. The meanings of the symbols are given in $\S$ \hyperref[Nomenclature]{Nomenclature}
with reference to Figure \ref{FigShapes}. See \cite{Whitebook1991,Lekner2007}.\vspace{0.3cm}}
\begin{tabular}{|l|l|}
\hline 
\textbf{Geometry of Cross Section} & \textbf{Volumetric Flow Rate ($\boldsymbol{Q}$)}\tabularnewline
\hline 
Ellipse & $Q=\frac{\partial p}{\partial z}\frac{\pi a^{3}b^{3}}{4\mu\left(a^{2}+b^{2}\right)}$\tabularnewline
\hline 
Rectangle & $Q=\frac{\partial p}{\partial z}\frac{4a^{3}b}{3\mu}\left[1-\frac{192a}{\pi^{5}b}\sum_{i=1,3,5,\ldots}^{\infty}\frac{\tanh\left(\frac{i\pi b}{2a}\right)}{i^{5}}\right]$\tabularnewline
\hline 
Equilateral Triangle & $Q=\frac{\partial p}{\partial z}\frac{\sqrt{3}\,a^{4}}{320\mu}$\tabularnewline
\hline 
Concentric Circular Annulus & $Q=\frac{\partial p}{\partial z}\frac{\pi}{8\mu}\left[a^{4}-b^{4}-\frac{\left(a^{2}-b^{2}\right)^{2}}{\ln(a/b)}\right]$\tabularnewline
\hline 
\end{tabular}\label{QGTable}
\par\end{centering}
\end{table}

\clearpage{}

\begin{table}
\centering{}\caption{The semi-major axes $a$, semi-minor axes $b$ and lengths $L$ of
the seven elliptical tubes of the 1D linear network of Figure \ref{NewtEllipseFig}
as well as the pressure drops $\Delta p$ across them. All numbers
are in SI units (i.e. meters and pascals).\vspace{0.0cm}}
\begin{tabular}{|l|l|}
\hline 
$a$ & 0.025, 0.021, 0.018, 0.013, 0.017, 0.026, 0.016\tabularnewline
\hline 
$b$ & 0.018, 0.019, 0.015, 0.011, 0.014, 0.018, 0.012\tabularnewline
\hline 
$L$ & 0.8, 0.6, 0.7, 0.9, 0.9, 0.5, 0.6\tabularnewline
\hline 
$\Delta p$ & 127.62, 116.04, 299.07, 1367.19, 495.96, 74.72, 519.40\tabularnewline
\hline 
\end{tabular}\label{TabNewtEllipseFig}
\end{table}

\newcommand{\CIF}     {\centering \includegraphics[width=3in]} %
\newcommand{\Vmin}    {\vspace{-0.2cm}} %
\newcommand{\Hs}      {\hspace{-0.5cm}} %
\begin{figure} %
\centering %
\subfigure[Conjugate Gradient.]%
{\begin{minipage}[b]{0.5\textwidth} \CIF {Illustrations/NewtEllipseFig1}
\end{minipage}} \Hs \subfigure[Nelder-Mead.]%
{\begin{minipage}[b]{0.5\textwidth} \CIF {Illustrations/NewtEllipseFig2}
\end{minipage}} \Vmin
\centering %
\subfigure[Quasi-Newton.]%
{\begin{minipage}[b]{0.5\textwidth} \CIF {Illustrations/NewtEllipseFig3}
\end{minipage}} \Hs \subfigure[Stochastic Global.]%
{\begin{minipage}[b]{0.5\textwidth} \CIF {Illustrations/NewtEllipseFig4}
\end{minipage}} %
\caption{Comparison between the analytical solution and the solutions obtained from the indicated global optimization algorithms which are based on the energy minimization principle for the flow of a Newtonian fluid with $\mu=0.05$ Pa.s. The computations were carried out using the 1D linear elliptical network of Table \ref{TabNewtEllipseFig} with inlet and outlet pressures of 3000~Pa and 0~Pa respectively. The volumetric flow rate through the network is 0.00024061 m$^3$.s$^{-1}$. In all four sub-figures, the vertical axis represents the network axial pressure in Pa while the horizontal axis represents the network axial coordinate in m. \label{NewtEllipseFig}}
\end{figure}

\clearpage{}

\begin{table}
\centering{}\caption{The semi-lengths $a$, semi-widths $b$ and lengths $L$ of the eight
rectangular tubes of the 1D linear network of Figure \ref{NewtRectFig}
as well as the pressure drops $\Delta p$ across them. All numbers
are in SI units (i.e. meters and pascals).\vspace{0.0cm}}
\begin{tabular}{|l|l|}
\hline 
$a$ & 0.19, 0.17, 0.21, 0.20, 0.22, 0.16, 0.18, 0.23\tabularnewline
\hline 
$b$ & 0.12, 0.15, 0.13, 0.14, 0.10, 0.15, 0.16, 0.09\tabularnewline
\hline 
$L$ & 1.1, 0.6, 0.8, 0.6, 1.1, 0.7, 1.3, 0.8\tabularnewline
\hline 
$\Delta p$ & 425.29, 169.92, 217.50, 148.25, 539.86, 222.61, 288.39, 488.19\tabularnewline
\hline 
\end{tabular}\label{TabNewtRectFig}
\end{table}

\begin{figure} %
\centering %
\subfigure[Conjugate Gradient.]%
{\begin{minipage}[b]{0.5\textwidth} \CIF {Illustrations/NewtRectFig1}
\end{minipage}} \Hs \subfigure[Nelder-Mead.]%
{\begin{minipage}[b]{0.5\textwidth} \CIF {Illustrations/NewtRectFig2}
\end{minipage}} \Vmin
\centering %
\subfigure[Quasi-Newton.]%
{\begin{minipage}[b]{0.5\textwidth} \CIF {Illustrations/NewtRectFig3}
\end{minipage}} \Hs \subfigure[Stochastic Global.]%
{\begin{minipage}[b]{0.5\textwidth} \CIF {Illustrations/NewtRectFig4}
\end{minipage}} %
\caption{Comparison between the analytical solution and the solutions obtained from the indicated global optimization algorithms which are based on the energy minimization principle for the flow of a Newtonian fluid with $\mu=0.13$ Pa.s. The computations were carried out using the 1D linear rectangular network of Table \ref{TabNewtRectFig} with inlet and outlet pressures of 2500~Pa and 0~Pa respectively. The volumetric flow rate through the network is 0.790778 m$^3$.s$^{-1}$. In all four sub-figures, the vertical axis represents the network axial pressure in Pa while the horizontal axis represents the network axial coordinate in m. \label{NewtRectFig}}
\end{figure}

\clearpage{}

\begin{table}
\centering{}\caption{The side lengths $a$ and lengths $L$ of the eight equilateral triangular
tubes of the 1D linear network of Figure \ref{NewtTriangFig} as well
as the pressure drops $\Delta p$ across them. All numbers are in
SI units (i.e. meters and pascals).\vspace{0.0cm}}
\begin{tabular}{|l|l|}
\hline 
$a$ & 0.019, 0.015, 0.021, 0.014, 0.022, 0.015, 0.018, 0.009\tabularnewline
\hline 
$L$ & 0.120, 0.070, 0.080, 0.060, 0.080, 0.060, 0.110, 0.070\tabularnewline
\hline 
$\Delta p$ & 78.83, 118.38, 35.22, 133.72, 29.24, 101.47, 89.71, 913.43\tabularnewline
\hline 
\end{tabular}\label{TabNewtTriangFig}
\end{table}

\begin{figure} %
\centering %
\subfigure[Conjugate Gradient.]%
{\begin{minipage}[b]{0.5\textwidth} \CIF {Illustrations/NewtTriangFig1}
\end{minipage}} \Hs \subfigure[Nelder-Mead.]%
{\begin{minipage}[b]{0.5\textwidth} \CIF {Illustrations/NewtTriangFig2}
\end{minipage}} \Vmin
\centering %
\subfigure[Quasi-Newton.]%
{\begin{minipage}[b]{0.5\textwidth} \CIF {Illustrations/NewtTriangFig3}
\end{minipage}} \Hs \subfigure[Stochastic Global.]%
{\begin{minipage}[b]{0.5\textwidth} \CIF {Illustrations/NewtTriangFig4}
\end{minipage}} %
\caption{Comparison between the analytical solution and the solutions obtained from the indicated global optimization algorithms which are based on the energy minimization principle for the flow of a Newtonian fluid with $\mu=0.1$ Pa.s. The computations were carried out using the 1D linear  triangular network of Table \ref{TabNewtTriangFig} with inlet and outlet pressures of 1500~Pa and 0~Pa respectively. The volumetric flow rate through the network is 4.634$\times 10^{-6}$ m$^3$.s$^{-1}$. In all four sub-figures, the vertical axis represents the network axial pressure in Pa while the horizontal axis represents the network axial coordinate in m. \label{NewtTriangFig}}
\end{figure}

\clearpage{}

\begin{table}
\centering{}\caption{The outer radii $a$, inner radii $b$ and lengths $L$ of the seven
concentric circular annular tubes of the 1D linear network of Figure
\ref{NewtAnnFig} as well as the pressure drops $\Delta p$ across
them. All numbers are in SI units (i.e. meters and pascals).\vspace{0.0cm}}
\begin{tabular}{|l|l|}
\hline 
$a$ & 0.050, 0.044, 0.052, 0.041, 0.056, 0.040, 0.051\tabularnewline
\hline 
$b$ & 0.045, 0.033, 0.039, 0.035, 0.041, 0.030, 0.043\tabularnewline
\hline 
$L$ & 0.61, 0.96, 0.88, 1.01, 0.77, 1.12, 0.95\tabularnewline
\hline 
$\Delta p$ & 623.80, 113.62, 53.39, 746.97, 28.52, 194.07, 239.63\tabularnewline
\hline 
\end{tabular}\label{TabNewtAnnFig}
\end{table}

\begin{figure}
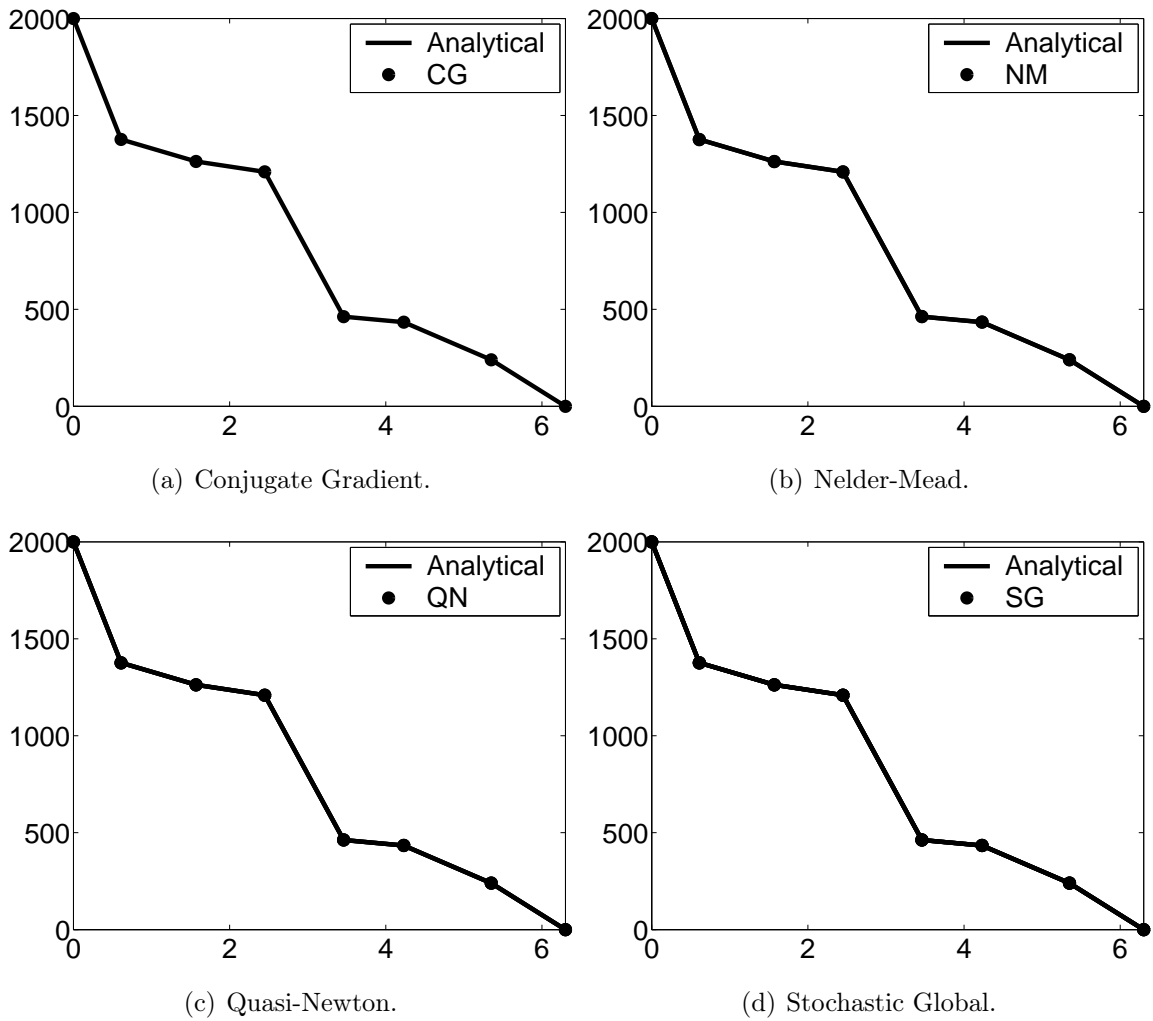
 %
\centering %
\subfigure[Conjugate Gradient.]%
{\begin{minipage}[b]{0.5\textwidth} \CIF {Illustrations/NewtAnnFig1}
\end{minipage}} \Hs \subfigure[Nelder-Mead.]%
{\begin{minipage}[b]{0.5\textwidth} \CIF {Illustrations/NewtAnnFig2}
\end{minipage}} \Vmin
\centering %
\subfigure[Quasi-Newton.]%
{\begin{minipage}[b]{0.5\textwidth} \CIF {Illustrations/NewtAnnFig3}
\end{minipage}} \Hs \subfigure[Stochastic Global.]%
{\begin{minipage}[b]{0.5\textwidth} \CIF {Illustrations/NewtAnnFig4}
\end{minipage}} %
\caption{Comparison between the analytical solution and the solutions obtained from the indicated global optimization algorithms which are based on the energy minimization principle for the flow of a Newtonian fluid with $\mu=0.15$ Pa.s. The computations were carried out using the 1D linear annular network of Table \ref{TabNewtAnnFig} with inlet and outlet pressures of 2000~Pa and 0~Pa respectively. The volumetric flow rate through the network is 2.120$\times 10^{-5}$ m$^3$.s$^{-1}$. In all four sub-figures, the vertical axis represents the network axial pressure in Pa while the horizontal axis represents the network axial coordinate in m. \label{NewtAnnFig}}
\end{figure}

\clearpage{}

\section{\label{secNonNewtonian}Non-Newtonian Flow in Elliptical Geometry}

In \cite{SochiOptMulVar2014} we demonstrated energy minimization
in the flow of generalized Newtonian fluids (which include Newtonian
and 6 non-Newtonian fluid types) through tubes and networks of interconnected
tubes of circular cross sections, while in $\S$ \ref{secNewtonianGeometries}
of the present paper we demonstrated energy minimization in the flow
of Newtonian fluids through various cross sectional geometries (namely
elliptical, rectangular, equilateral triangular, and concentric circular
annular). In the present section we combine these two variations (i.e.
the variation of fluid type and the variation of cross sectional geometry)
by demonstrating energy minimization in the flow of non-Newtonian
fluids (namely power law, Ellis and Ree-Eyring fluids) through tubes
of elliptical cross sectional geometry.

In this regard we use the analytical expressions of Table \ref{QETable}
(as given in \cite{SochiEllipsePL2025,SochiEllipseEllis2025,SochiEllipseRE2025})
for the volumetric flow rate of power law, Ellis and Ree-Eyring fluids
through tubes and networks of interconnected tubes of elliptical cross
sectional geometry where the coordinate and parametric setting of
the elliptical geometry with regard to these expressions is illustrated
in Figure \ref{FigShapes}.

We conducted thorough investigations and computations of non-Newtonian
fluid flow through elliptical tubes and networks (i.e. of interconnected
elliptical tubes) of various dimensionality (i.e. 1D, 2D and 3D) and
topology (such as linear, fractals, and irregulars based on cubic
and orthorhombic lattices) using the three aforementioned non-Newtonian
fluid types (i.e. power law, Ellis and Ree-Eyring) of various characteristics
and parameters. All these investigations and computations lead to
the definite conclusion that energy minimization holds in all these
cases regardless of the type of elliptical vessel (i.e. whether single
tube or network of interconnected tubes), regardless of the dimensionality
and topology of network, and regardless of the characteristics and
parameters of fluid.

\begin{figure}
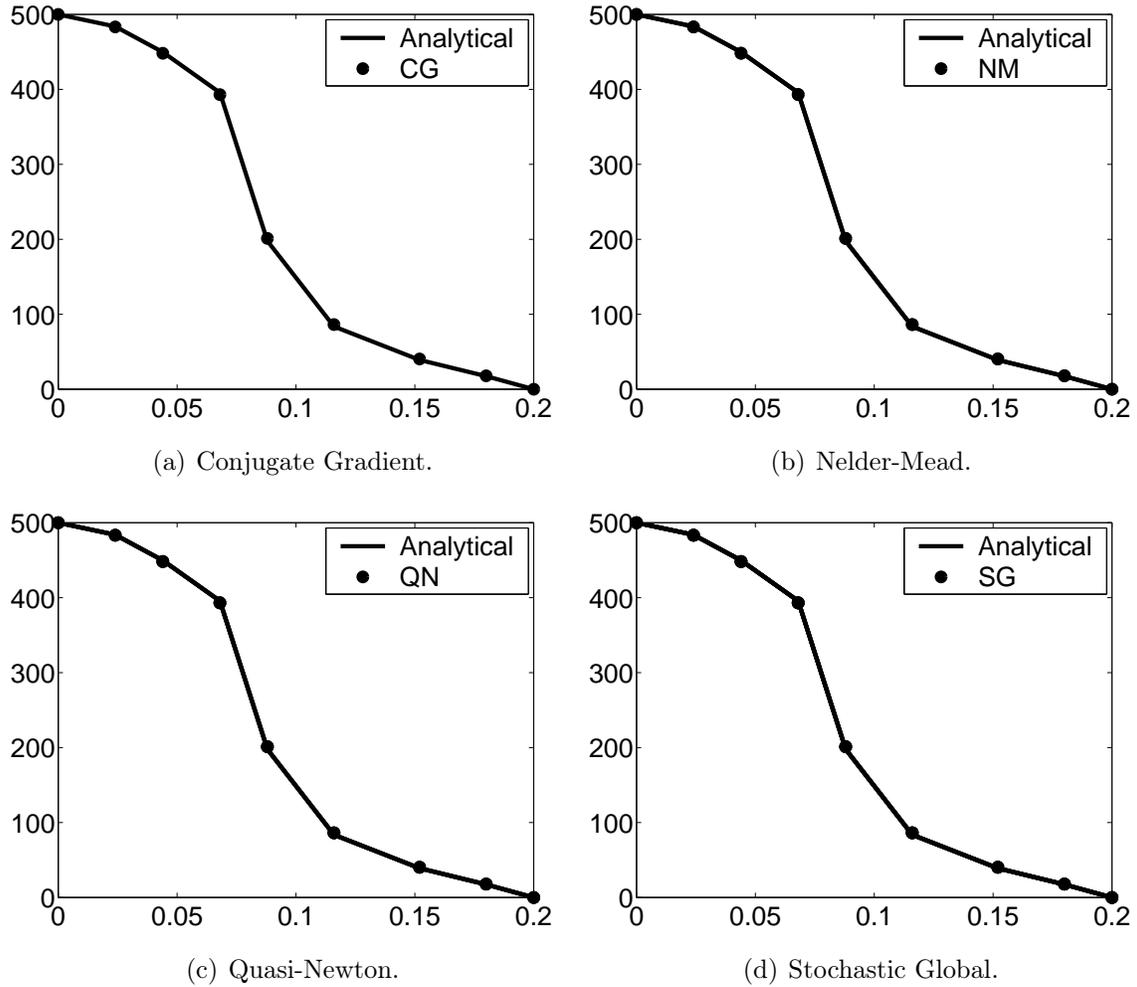
 %
\centering %
\subfigure[Conjugate Gradient.]%
{\begin{minipage}[b]{0.5\textwidth} \CIF {Illustrations/EllisEllipseFig1}
\end{minipage}} \Hs \subfigure[Nelder-Mead.]%
{\begin{minipage}[b]{0.5\textwidth} \CIF {Illustrations/EllisEllipseFig2}
\end{minipage}} \Vmin
\centering %
\subfigure[Quasi-Newton.]%
{\begin{minipage}[b]{0.5\textwidth} \CIF {Illustrations/EllisEllipseFig3}
\end{minipage}} \Hs \subfigure[Stochastic Global.]%
{\begin{minipage}[b]{0.5\textwidth} \CIF {Illustrations/EllisEllipseFig4}
\end{minipage}} %
\caption{Comparison between the analytical solution and the solutions obtained from the indicated global optimization algorithms which are based on the energy minimization principle for the flow of an Ellis fluid with $\alpha=2.4$, $\mu_e=0.18$ Pa.s and $\tau_h=40$ Pa. The computations were carried out using the 1D linear elliptical network of Table \ref{TabEllisEllipseFig} with inlet and outlet pressures of 500~Pa and 0~Pa respectively. The volumetric flow rate through the network is 1.092$\times 10^{-6}$ m$^3$.s$^{-1}$. In all four sub-figures, the vertical axis represents the network axial pressure in Pa while the horizontal axis represents the network axial coordinate in m. \label{EllisEllipseFig}}
\end{figure}

\begin{figure}
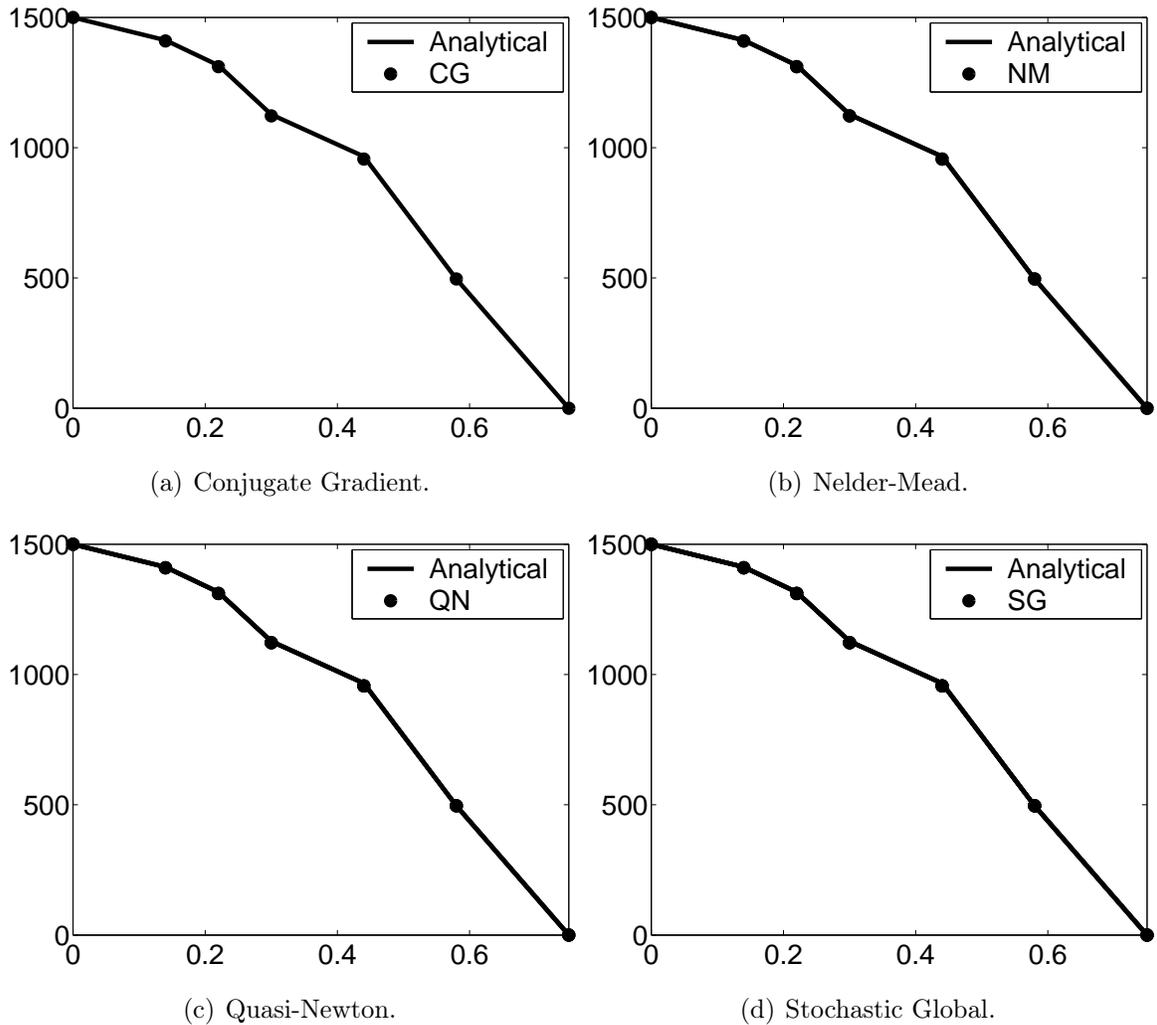
 %
\centering %
\subfigure[Conjugate Gradient.]%
{\begin{minipage}[b]{0.5\textwidth} \CIF {Illustrations/ReeEllipseFig1}
\end{minipage}} \Hs \subfigure[Nelder-Mead.]%
{\begin{minipage}[b]{0.5\textwidth} \CIF {Illustrations/ReeEllipseFig2}
\end{minipage}} \Vmin
\centering %
\subfigure[Quasi-Newton.]%
{\begin{minipage}[b]{0.5\textwidth} \CIF {Illustrations/ReeEllipseFig3}
\end{minipage}} \Hs \subfigure[Stochastic Global.]%
{\begin{minipage}[b]{0.5\textwidth} \CIF {Illustrations/ReeEllipseFig4}
\end{minipage}} %
\caption{Comparison between the analytical solution and the solutions obtained from the global optimization algorithms which are based on the energy minimization principle for the flow of a Ree-Eyring fluid with  $\mu_0=0.018$ Pa.s and $\tau_c=30$ Pa. The computations were carried out using the 1D linear elliptical network of Table \ref{TabReeEllipseFig} with inlet and outlet pressures of 1500~Pa and 0~Pa respectively. The volumetric flow rate through the network is 0.0034445 m$^3$.s$^{-1}$. In all four sub-figures, the vertical axis represents the network axial pressure in Pa while the horizontal axis represents the network axial coordinate in m. \label{ReeEllipseFig}}
\end{figure}

A sample of these investigations and computations for some 1D linear
networks (see Tables \ref{TabPLEllipseFig}-\ref{TabReeEllipseFig})
are given in Figures \ref{PLEllipseFig}-\ref{ReeEllipseFig}, i.e.
Figure \ref{PLEllipseFig} for the flow of a power law fluid through
elliptical geometry, Figure \ref{EllisEllipseFig} for the flow of
an Ellis fluid through elliptical geometry, and Figure \ref{ReeEllipseFig}
for the flow of a Ree-Eyring fluid through elliptical geometry. The
use of 1D linear networks in this sample demonstration is because
of the ease of displaying and clarity of showing and interpreting
the results of this type of networks which can be easily displayed
on 2D plots (as seen in Figures \ref{PLEllipseFig}-\ref{ReeEllipseFig})
and clearly analyzed and interpreted. 

As we see, this representative sample shows that the solutions that
we obtained from the employment of the principle of energy minimization
are virtually identical to the Poiseuille-type analytical solutions
(as outlined in $\S$ \ref{secBackground}). So, we can conclude with
certainty that the principle of energy minimization holds in general
(i.e. independent of the type of vessel, the dimensionality and topology
of network, and the type of generalized Newtonian fluid).

\begin{table}[t]
\caption{The three types of generalized Newtonian fluid models and the volumetric
rates $Q$ for their flow through tubes of elliptical cross sectional
geometry. The meanings of the symbols are given in $\S$ \hyperref[Nomenclature]{Nomenclature}
with reference to Figure \ref{FigShapes} $\Big($noting that $A=\frac{\partial p}{\partial z}\frac{1}{a^{2}+b^{2}}\Big)$.
See \cite{SochiEllipsePL2025,SochiEllipseEllis2025,SochiEllipseRE2025}.\vspace{0.3cm}}
\begin{tabular}{|l|l|}
\hline 
\textbf{Fluid Type} & \textbf{Volumetric Flow Rate ($\boldsymbol{Q}$)}\tabularnewline
\hline 
Power Law & $Q=\frac{4\,\arctan\left(\frac{\tan(\pi/2)}{\sqrt{1-e^{2}}}\right)}{\sqrt{1-e^{2}}}\left(\frac{\partial p}{\partial z}\,\frac{b^{2}}{k(a^{2}+b^{2})}\right)^{1/n}\frac{n}{n+1}\left[\frac{nb^{(3n+1)/n}}{(3n+1)\left(1-e^{2}\right)^{(n+1)/(2n)}}-\frac{a^{(n+1)/n}b^{2}}{2}\right]$\tabularnewline
\hline 
\multirow{2}{*}{Ellis} & \multicolumn{1}{l|}{$Q=\left\{ Ab^{4}\left[\frac{b^{2}}{2\left(1-e^{2}\right)}-a^{2}\right]+\frac{4A^{\alpha}b^{2\alpha+2}}{\tau_{h}^{\alpha-1}(\alpha+1)}\left[\frac{b^{\alpha+1}}{\left(1-e^{2}\right)^{(\alpha+1)/2}(\alpha+3)}-\frac{a^{\alpha+1}}{2}\right]\right\} $}\tabularnewline
 & $\hspace{10.4cm}\frac{\arctan\left(\frac{\tan(\pi/2)}{\sqrt{1-e^{2}}}\right)}{\mu_{e}\sqrt{1-e^{2}}}$\tabularnewline
\hline 
\multirow{2}{*}{Ree-Eyring} & \multicolumn{1}{l|}{$Q=\frac{4\tau_{c}^{4}\sqrt{1-e^{2}}}{A^{3}b^{6}\mu_{0}}\Bigg[\frac{Ab^{3}}{\tau_{c}\sqrt{1-e^{2}}}\sinh\left(\frac{Ab^{3}}{\tau_{c}\sqrt{1-e^{2}}}\right)-\cosh\left(\frac{Ab^{3}}{\tau_{c}\sqrt{1-e^{2}}}\right)$}\tabularnewline
 & $\hspace{5cm}-\frac{A^{2}b^{6}}{2\tau_{c}^{2}\left(1-e^{2}\right)}\cosh\left(\frac{Ab^{2}a}{\tau_{c}}\right)+1\Bigg]\arctan\left(\frac{\tan(\pi/2)}{\sqrt{1-e^{2}}}\right)$\tabularnewline
\hline 
\end{tabular}\label{QETable}
\end{table}

\clearpage{}

\begin{table}
\centering{}\caption{The semi-major axes $a$, semi-minor axes $b$ and lengths $L$ of
the seven elliptical tubes of the 1D linear network of Figure \ref{PLEllipseFig}
as well as the pressure drops $\Delta p$ across them. All numbers
are in SI units (i.e. meters and pascals).\vspace{0.0cm}}
\begin{tabular}{|l|l|}
\hline 
$a$ & 0.017, 0.023, 0.019, 0.021, 0.018, 0.024, 0.022\tabularnewline
\hline 
$b$ & 0.013, 0.017, 0.016, 0.016, 0.014, 0.019, 0.015\tabularnewline
\hline 
$L$ & 0.7, 0.8, 0.9, 0.9, 1.0, 0.9, 0.8\tabularnewline
\hline 
$\Delta p$ & 599.22, 185.48, 365.89, 294.12, 631.30, 144.88, 279.11\tabularnewline
\hline 
\end{tabular}\label{TabPLEllipseFig}
\end{table}

\begin{figure} %
\centering %
\subfigure[Conjugate Gradient.]%
{\begin{minipage}[b]{0.5\textwidth} \CIF {Illustrations/PLEllipseFig1}
\end{minipage}} \Hs \subfigure[Nelder-Mead.]%
{\begin{minipage}[b]{0.5\textwidth} \CIF {Illustrations/PLEllipseFig2}
\end{minipage}} \Vmin
\centering %
\subfigure[Quasi-Newton.]%
{\begin{minipage}[b]{0.5\textwidth} \CIF {Illustrations/PLEllipseFig3}
\end{minipage}} \Hs \subfigure[Stochastic Global.]%
{\begin{minipage}[b]{0.5\textwidth} \CIF {Illustrations/PLEllipseFig4}
\end{minipage}} %
\caption{Comparison between the analytical solution and the solutions obtained from the indicated global optimization algorithms which are based on the energy minimization principle for the flow of a power law fluid with $n=1.2$ and $k=0.07$ Pa.s$^n$. The computations were carried out using the 1D linear elliptical network of Table \ref{TabPLEllipseFig} with inlet and outlet pressures of 2500~Pa and 0~Pa respectively. The volumetric flow rate through the network is 0.00011459  m$^3$.s$^{-1}$. In all four sub-figures, the vertical axis represents the network axial pressure in Pa while the horizontal axis represents the network axial coordinate in m. \label{PLEllipseFig}}
\end{figure}

\clearpage{}

\begin{table}
\centering{}\caption{The semi-major axes $a$, semi-minor axes $b$ and lengths $L$ of
the eight elliptical tubes of the 1D linear network of Figure \ref{EllisEllipseFig}
as well as the pressure drops $\Delta p$ across them. All numbers
are in SI units (i.e. meters and pascals).\vspace{0.0cm}}
\begin{tabular}{|l|l|}
\hline 
$a$ & 0.0060, 0.0050, 0.0044, 0.0028, 0.0038, 0.0049, 0.0057, 0.0051\tabularnewline
\hline 
$b$ & 0.0046, 0.0035, 0.0034, 0.0024, 0.0029, 0.0041, 0.0045, 0.0047\tabularnewline
\hline 
$L$ & 0.024, 0.020, 0.024, 0.020, 0.028, 0.036, 0.028, 0.020\tabularnewline
\hline 
$\Delta p$ & 16.20, 34.16, 54.02, 197.90, 114.08, 44.66, 21.70, 17.28\tabularnewline
\hline 
\end{tabular}\label{TabEllisEllipseFig}
\end{table}

\clearpage{}

\begin{table}
\centering{}\caption{The semi-major axes $a$, semi-minor axes $b$ and lengths $L$ of
the six elliptical tubes of the 1D linear network of Figure \ref{ReeEllipseFig}
as well as the pressure drops $\Delta p$ across them. All numbers
are in SI units (i.e. meters and pascals).\vspace{0.0cm}}
\begin{tabular}{|l|l|}
\hline 
$a$ & 0.024, 0.020, 0.017, 0.022, 0.015, 0.018\tabularnewline
\hline 
$b$ & 0.021, 0.018, 0.015, 0.017, 0.014, 0.013\tabularnewline
\hline 
$L$ & 0.14, 0.08, 0.08, 0.14, 0.14, 0.17\tabularnewline
\hline 
$\Delta p$ & 87.34, 96.60, 188.52, 161.50, 469.82, 496.22\tabularnewline
\hline 
\end{tabular}\label{TabReeEllipseFig}
\end{table}

\clearpage{}

\section{\label{secConclusions}Conclusions}

We outline in the following points the main achievements and conclusions
of the present paper:
\begin{enumerate}
\item We continued our previous investigations about energy minimization
in the flow of generalized Newtonian fluids through tubes and networks
of interconnected tubes of various geometries. In the present study
we extended and generalized the previous investigations by including
non-circular and non-simply-connected geometries of tubes and networks
of such tubes where we extended the investigation of this principle
to the flow of Newtonian fluids through some of the common non-circular
and non-simply-connected geometries (namely elliptical, rectangular,
equilateral triangular and concentric circular annular geometries).
We also extended this principle to the flow of three non-Newtonian
fluid types (namely power law, Ellis and Ree-Eyring) through tubes
and networks of interconnected tubes of elliptical cross sections.
\item All the results obtained in the current study support the generalization
of the energy minimization as a fundamental principle that underlies
the flow phenomena in the Newtonian and non-Newtonian fluid dynamics
systems. The study also adds more tools to the subjects of Computational
Fluid Dynamics (CFD) as these computational methods, which are based
on energy minimization, can be used for finding the pressure and flow
rate fields in tubes and networks and could have some advantages in
some cases over the traditional methods (as indicated and concluded
in our previous investigations).
\item The main conclusion of this investigation is that energy minimization
in the flow of generalized Newtonian fluids through tubes and networks
of interconnected tubes is general, i.e. it is independent of the
type and geometry of the flow vessels and their topologies and it
is independent of the type of fluids and their rheological properties.
In fact, we may dare to say that energy minimization is a valid principle
in all fluid dynamics systems regardless of the properties of vessels,
fluids and flow conditions (i.e. within the restrictions and constraints
of the particular fluid dynamics system and its physical conditions).
\item As well as its obvious theoretical value, the present investigation
and the obtained results have practical values for the Computational
Fluid Dynamics (as indicated already).
\item The relevance and usefulness of the present investigation and the
obtained results extend beyond the subject of fluid flow through tubes
and networks of interconnected tubes to include, for instance, the
flow of fluids through porous media and the flow of electric currents
through certain types of electrical networks such as networks of interconnected
ohmic conductors.
\end{enumerate}
\phantomsection 
\addcontentsline{toc}{section}{References}\bibliographystyle{unsrt}
\bibliography{Bibl}
\phantomsection \addcontentsline{toc}{section}{Nomenclature}

\section*{\label{Nomenclature}Nomenclature}

\noindent %
\begin{longtable}[l]{ll}
1D, 2D, 3D & one dimensional, two dimensional, three dimensional\tabularnewline
\addlinespace[0.05cm]
$a$ & length of side of equilateral triangle (m)\tabularnewline
\addlinespace[0.05cm]
$a,b$ & semi-major and semi-minor axes of ellipse (m)\tabularnewline
\addlinespace[0.05cm]
$a,b$ & semi-length and semi-width of rectangle (m)\tabularnewline
\addlinespace[0.05cm]
$a,b$ & outer and inner radii of concentric circular annulus (m)\tabularnewline
\addlinespace[0.05cm]
$e$ & eccentricity of ellipse (~)\tabularnewline
\addlinespace[0.05cm]
Eq., Eqs. & Equation, Equations\tabularnewline
\addlinespace[0.05cm]
$I$ & time rate of energy consumption for fluid transport (J.s$^{-1}$)\tabularnewline
\addlinespace[0.05cm]
$I_{t}$ & time rate of total energy consumption for fluid transport (J.s$^{-1}$)\tabularnewline
\addlinespace[0.05cm]
$k$ & viscosity coefficient in the power law model (Pa.s$^{n}$)\tabularnewline
\addlinespace[0.05cm]
$L$ & tube length (m)\tabularnewline
\addlinespace[0.05cm]
$m$ & number of discrete/discretized elements in fluid conducting device
(~)\tabularnewline
\addlinespace[0.05cm]
$n$ & flow behavior index in the power law model (~)\tabularnewline
\addlinespace[0.05cm]
$N$ & number of nodal junctions in fluid conducting device (~)\tabularnewline
\addlinespace[0.05cm]
$p$ & pressure (Pa)\tabularnewline
\addlinespace[0.05cm]
$Q$ & volumetric flow rate (m$^{3}$.s$^{-1}$)\tabularnewline
\addlinespace[0.05cm]
$x,y,z$ & coordinate variables (usually spatial coordinates)\tabularnewline
\addlinespace[0.05cm]
 & \tabularnewline
\addlinespace[0.05cm]
$\alpha$ & indicial parameter in Ellis model ( )\tabularnewline
\addlinespace[0.05cm]
$\Delta p$ & pressure drop across flow conduit (Pa)\tabularnewline
\addlinespace[0.05cm]
$\mu$ & Newtonian viscosity (Pa.s)\tabularnewline
\addlinespace[0.05cm]
$\mu_{0}$ & low-shear viscosity in Ree-Eyring model (Pa.s)\tabularnewline
\addlinespace[0.05cm]
$\mu_{e}$ & low-shear viscosity in Ellis model (Pa.s)\tabularnewline
\addlinespace[0.05cm]
$\tau_{c}$ & characteristic shear stress in Ree-Eyring model (Pa)\tabularnewline
\addlinespace[0.05cm]
$\tau_{h}$ & shear stress when viscosity equals $\frac{\mu_{e}}{2}$ in Ellis model
(Pa)\tabularnewline
\addlinespace[0.05cm]
\end{longtable}
\end{document}